\documentstyle[epsf]{elsart}

\begin{document}

\begin{frontmatter}
    
\title{Nuclear and Neutron Matter Calculations with Different Model 
Spaces}

\author{L.\ Engvik, E.\ Osnes}
\address{Department of Physics, University of Oslo, N-0316 Oslo, 
Norway}
\author{M.\ Hjorth-Jensen}
\address{Nordita, Blegdamsvej 17, DK-2100 K\o benhavn \O, Denmark}
\author{T.T.S.\ Kuo}
\address{Department of Physics, State University of New York
at Stony Brook, NY 11794, USA }
\maketitle

\begin{abstract}
In this work we investigate the so-called model-space 
Brueckner-Hartree-Fock 
(MBHF) approach for nuclear matter as well as for neutron matter 
and the extension of this which  
includes  the particle-particle and hole-hole (PPHH) diagrams.
A central ingredient in the model-space approach for nuclear matter
is the boundary momentum $k_M$ 
beyond which the single-particle potential energy is set equal to zero.
This is also the boundary of the model space within which 
the PPHH diagrams are calculated.
It has been rather uncertain which value  should be used for $k_M$.
We have carried out model-space nuclear matter and neutron matter 
calculations with and without PPHH diagrams for various 
choices of $k_M$
and using several modern nucleon-nucleon potentials.
Our results exhibit a saturation region where  the 
nuclear and neutron matter 
matter energies  are quite stable  as $k_M$ varies.
The location of this region may serve to determine an "optimum" choice
for $k_M$.
However, we find that the strength of the tensor force has a 
significant influence on binding energy variation with $k_M$.
The implications for nuclear and neutron matter calculations are 
discussed.
\end{abstract}
\end{frontmatter}

\section{Introduction}

A common problem to  non-relativistic nuclear matter 
calculations has been the
simultaneous reproduction of both the binding  energy per nucleon
($BE/A=16\pm 1$ MeV) and the saturation density
$\rho_0=0.17$ fm$^{-3}$. All modern realistic calculations of
nuclear matter which employ two-body interactions such as
Brueckner-like approaches  \cite{rpd89,rdp91,dm92,km83,shk87,mahaux85},
hyper-netted-chain \cite{hnc} or coupled cluster \cite{cs78} approaches
are not able to reproduce in a satisfactory way 
the above nuclear matter data. If for a given interaction
one is able to predict the saturation density, then the binding energy
is underestimated. Similarly, if one is able to reproduce the
binding energy of nuclear matter, then the saturation density is
too high. This results in the so-called Coester band 
(see e.g. Fig.\ \ref{fig:mbhfsat} below), 
where the saturation
point for a given nucleon-nucleon (NN) interaction
is correlated with the strength of the tensor force in the 
NN interaction.

In the past years relativistic approaches to nuclear matter 
have also been developed. 
For example, Brockmann and Machleidt \cite{mac89,bm90}
have recently shown, starting from a realistic NN 
interaction,
that by performing a relativistic Dirac-Brueckner-Hartree-Fock (DBHF)
nuclear matter calculation, one is able to meet the empirical nuclear
matter data.
Another line of approach is represented by the relativistic
mean field scheme of Serot and Walecka \cite{sw86}. However, the
present relativistic approaches neglect the coupling to the
negative energy solutions, and moreover
relativistic approaches lead to a very small depletion of the
Fermi sea \cite{jm88}. 
This depletion is not consistent with recent $(e,e'p)$ 
experiments. 

Therefore, we believe that it is still worth to properly investigate
non-relativistic schemes based on extensions of the 
Brueckner-Hartree-Fock (BHF) approach.
One of the main problems with Brueckner theory is that particle states 
are treated differently from hole states.
In the standard BHF approach one employs a discontinuous 
single particle (sp) spectrum, with a 
self-consistent BHF part for the nucleons below  and a 
free-particle part above the Fermi surface. 
This leads to an unphysically large discontinuity 
of about 60 MeV for the sp spectrum at the Fermi level.

Several methods have been proposed to overcome this difficulty. 
Mahaux and his collaborators\cite{mahaux85} used  the real part of the 
reaction matrix
to obtain a self-consistent sp spectrum which is continuous for all 
momenta.
In the model-space Brueckner-Hartree-Fock (MBHF) approach proposed by 
Ma and Kuo\cite{km83}
a continuous sp spectrum is obtained within a chosen model space
(defined for momenta $k\leq k_M$), and the discontinuity at the 
model space boundary  $k_M$ is 
rather small for proper values of $k_M$.
Although both methods yield more binding energy than 
the standard BHF approach, the nuclear matter 
empirical data are not reproduced. 
However, 
by including PPHH diagrams in the model space approach 
Song et al. \cite{shk87} and Jiang et al.
\cite{jia88} have demonstrated that the saturation properties  
can be improved.   
Comparing with the standard hole-line expansion
in BHF  calculations, it is remarkable
that the PPHH diagrams calculation can simultaneously increase
the nuclear-matter binding energy and lower the
saturation density $\rho _0$. 

A central question concerning the above PPHH-diagram nuclear matter
approach is how to choose the boundary momentum  $k_M$.
In the past \cite{shk87,jia88}, one usually took a "reasonable"
value for the size of the model space such as $k_M$ = 3.2 fm$^{-1}$ 
and proceeded to evaluate the nuclear matter PPHH diagrams. 
To our knowledge, PPHH-diagram
nuclear matter calculations with other choices of $k_M$  have not been
investigated in a systematic way. Are there certain criteria which may
help determine the boundary $k_M$?
Is there a saturation behavior in the sense that
there exists a region where the results of PPHH-diagram nuclear
matter calculations are insensitive to the choice of $k_M$? 
The aim of this
work is to study these questions.
The answers to these questions may help  confirm, or disconfirm, the
rather encouraging results of the PPHH-diagram calculations reported 
earlier.

The reader should note that another way of summing  PPHH diagrams has
been developed by Dickhoff, Polls and Ramos, see e.g., Refs.\ 
\cite{rpd89,rdp91,dm92}. There a self-consistent 
Green's function (SCGF) approach
is used, where one solves the Dyson equation in order to
get the sp Green's function. Although the sp
energies are kept real in the self-consistency scheme, 
as in our case as well,
the two methods differ in the treatment of the poles in the 
in-medium scattering matrix. In the model-space
approach, see the discussion in the two next sections
and Ref.\ \cite{shk87}, there are, by 
construction no poles 
at negative and positive starting energies 
in the effective interaction defined
by the model-space reaction matrix.  The poles 
from the retarded Green's function are circumvented by a contour 
integration \cite{shk87}.
One avoids thereby a numerically tedious evaluation
of the two-body Green's function and effective interaction, as done
in Refs.\  \cite{rpd89,rdp91,dm92}.
However, the SCGF method of Dickhoff, Polls and Ramos 
\cite{rpd89,rdp91,dm92} allows one in a direct way to study 
other properties than just the binding energy. The one-body Green's
functions can be used e.g., in the study of momentum distributions and
spectral functions. 

Finally,  
our scheme to obtain an effective interaction appropriate for 
nuclear matter
starts with a free nucleon-nucleon  interaction $V$ which is
appropriate for nuclear physics at low and intermediate energies. 
At present there are several potentials available. 
The most recent versions of Machleidt and co-workers
\cite{cdbonn}, the Nijmegen group \cite{nim} and the Argonne
group \cite{v18} have a $\chi^2$ fit per datum close to $1$. 
The potential
model of Ref.\ \cite{cdbonn} is an extension of the one-boson-exchange
models of the Bonn group \cite{mac89}, where mesons like 
$\pi$, $\rho$, $\eta$, $\delta$, $\omega$ and the fictitious
$\sigma$ meson are included. In the charge-dependent version
of Ref.\ \cite{cdbonn}, the first five mesons have the same set
of parameters for all partial waves, whereas the parameters of
the $\sigma$ meson are allowed to vary. The recent Argonne potential
\cite{v18} is a charge-dependent version of the Argonne
$V14$ \cite{v14} potential. The Argonne potential models
are local potentials in coordinate space and include
a $\pi$-exchange plus parameterizations of the short-range
and intermediate range parts of the potential. The Nijmegen group   
\cite{nim} 
has constructed potentials based on meson exchange and models 
parameterized
in similar ways as the Argonne potentials.
Another important difference between e.g., the Bonn potentials
and the Argonne and Nijmegen potentials is the strength of the 
much debated
tensor force \cite{bm95}. Typically, the Bonn potentials have 
a smaller $D$-state admixture in the deuteron wave function
than the Argonne and Nijmegen potentials, as well as
other potential models. A smaller(larger) $D$-state
admixture in the ground state of the deuteron 
means that the tensor force is weaker(stronger).
The strength of the tensor force has important consequences 
in calculations of the binding energy for both
finite nuclei and infinite nuclear matter, see e.g., the discussion
in Ref.\ \cite{hko95}. A potential model  with a weak tensor force
tends to yield more attraction in a nuclear system than a 
potential with a strong tensor force.
The second  aim of this work, is therefore to compare various
nucleon-nucleon potentials, in order to study 
the role played by the tensor force.  
For this purpose we will work with 
 three older versions of the 
Bonn potentials defined in table A.1 of ref. \cite{mac89}. 
The only essential difference between these potentials is the 
relative strength of tensor force, which make the role of the 
tensor force more transparent in our investigation.
They are recognized by the labels  A, B and C with the former carrying 
the weakest tensor force. Such a set of potentials, fit to the same
set of scattering data, allow us therefore to study the role played 
by the nuclear tensor force   in a many-body approach like the present.
Moreover, these potentials are constructed in exactly the same
way, the only differences being the value of various meson
variables (coupling constants, masses and cutoff energies).
One may  therefore be able to ascribe possible differences to 
the role played by certain mesons, or certain components of the
NN interaction. 
 In addition we have also done extensive calculations using
the charge-dependent
version of the Bonn potential models (CD-Bonn), 
see Ref.\ \cite{cdbonn}, 
and two potentials of the Nijmegen group \cite{nim}, one
 meson-exchange potential (Nijm-OBE) and a local potential with 
a non-local contribution to the central force, i.e.\ the model
Nijm-I of Ref.\ \cite{nim}.
The reason for these choices is due to the fact that these
potential models represent the most recent fits to the scattering data.

The PPHH-diagram nuclear matter calculations are closely related to 
the MBHF calculations, both employing a model space ($k\leq k_M$).
In section \ref{sec:mbhf} we will present the MBHF method  and
the results obtained with it.
The basic method  presented in section \ref{sec:mbhf} is also used in 
the PPHH-diagram method discussed in section \ref{sec:pphh}.
The results of the PPHH-diagram calculations for nuclear and 
neutron matter are 
presented and discussed in sections\ \ref{sec:result} and 
\ref{sec:neutron},
respectively.
In section\ \ref{sec:concl} we summarize and conclude.
 
\section{Model Space Brueckner-Hartree-Fock Approach}
\label{sec:mbhf}
The basic formalism of the MBHF has been exposed elsewhere, see e.g.,
Refs.\ \cite{km83,shk87}. Here we will therefore 
only briefly sketch the essential 
ingredients of the method.

\subsection{The Model Space $G$-matrix and Single-particle Spectrum}
Following the conventional many-body approach, we rewrite the full
Hamiltonian $H=T+V$, $T$ being the kinetic energy
and $V$ the bare NN potential,
in an unperturbed part $H_0 =T+U$ and an interacting part $H_I = V-U$
such that
$H=T+V=H_0 + H_I$,
where we have introduced an auxiliary sp  potential $U$. If
$U$ is chosen such that $H_I$ be small, perturbative
many-body techniques can presumably be applied.
A serious obstacle to any perturbative treatment is the fact that the
bare NN potential $V$ is very large at small internucleon distances,
a fact which renders any 
perturbative approach highly prohibitive. To overcome
this problem, one introduces the reaction matrix $G$, given
by the solution of the integral equation
\begin{equation}
           G^M_{ijkl}(\omega)=\bar V_{ijkl}+\sum _{m,n}\bar V_{ijmn}
           \frac{Q^M(m,n)}{\omega-(\epsilon^M_m+ \epsilon^M_n)}
           G^M_{mnkl}(\omega),
           \label{eq:mbhf}
\end{equation}
with the model-space Pauli exclusion operator
\begin{equation}
      Q^M(m, n ) =\left\{\begin{array}{clc}
      1, &\min(k_m ,k_n ) > k_F&\mathrm{and}\\
      &\ \ \max(k_m ,k_n ) > k_M, &\\
      &&\\
      0,&\mbox{ otherwise}.&
			   \end{array}\right.
      \label{eq:pauliop}
\end{equation}
The model-space sp spectrum is defined self-consistently by
\begin{equation}
    \varepsilon^M_k = t_k + u^M_k,
    \label{eq:selfconx}
\end{equation}
where $t_k$ is the kinetic energy and the sp potential is
\begin{equation}
     u^M_k =\left\{    \begin{array}{ll}\\
     \sum_{h < k_F}
     \left\langle kh \right| G^M(\omega = 
     \varepsilon^M_k + \varepsilon^M_h )
     \left| kh \right\rangle , & k < k_M,  \\ \\
     0,& k > k_M.  \end{array}\right.
     \label{eq:selfcon}
\end{equation}
\begin{figure}[hbtp]
       \setlength{\unitlength}{1mm}
       \begin{picture}(80,100)
       \put(25,5){\epsfxsize=10cm \epsfbox{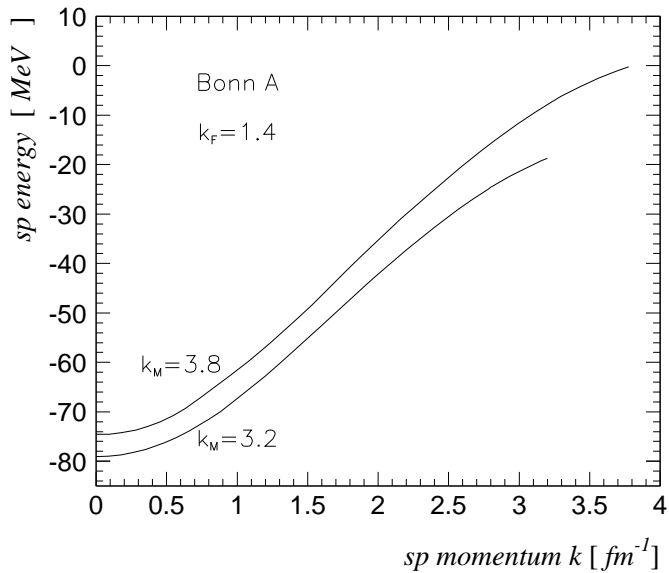}}
       \end{picture}
       \caption{The MBHF sp potential $u^M_k$ shown for  
                different values of $k_M$ 
                using the  Bonn A potential ($k_F=1.4$ fm$^{-1}$).}
       \label{fig:sp}
\end{figure}

The above MBHF approach is a generalization of the well-known BHF 
approach, the latter corresponding to a special case of the former 
with $k_M=k_F$ and commonly referred to as ``standard'' BHF.
For the BHF method, only the kinetic energy (free particle spectrum)
is included in the sp energy for  momenta greater than
$k_F$, while below $k_F$ the sp potential $u^M_k$ is included
as well. In the MBHF method, the corresponding boundary
is at $k_M$, beyond which one uses the free-particle spectrum.
The BHF and MBHF sp energies are constructed so that 
they are always real. 
To ensure this property, we need to require  that the MBHF spectrum 
$\epsilon ^M(k)$  
has a sp  potential 
$u^M_k$ which is always attractive. 
(Otherwise the propagator in
$G^M$ will have poles, leading to a complex $G^M$.) 
In Fig.\ \ref{fig:sp}, 
we show the
sp potential $u^M_k$ for some typical values of $k_M$.
Similar results are obtained using the other potentials. 
Generally speaking, when $k_M$ is small the sp potential is attractive.
As $k_M$ increases, the MBHF sp potential  becomes more repulsive. 
For $k_M~ =$ 3.8 fm$^{-1}$,  $\ u^M_k$  is still attractive
for small k, 
but becomes less attractive  for larger 
$k$ and is close to zero  at $k=k_M$. For $k_M~ >$ 3.8 fm$^{-1}$,
the sp  potential  becomes repulsive.
When this happens, we will have a crossing between the free 
particle spectrum and the MBHF spectrum. 
This will in turn give rise to a  sp potential
which is no longer real, 
unless
some $ad~hoc$ procedure is imposed such as suppressing its 
imaginary part.
To prevent such crossings, we have chosen 
$k_M~ \sim$ 3.8 fm$^{-1}$ as an
upper bound for $k_M$, which defines our model space.
This caveat ensures that the MBHF $G$-matrix
is always real. 

The MBHF approach is basically an intermediate scheme, 
in the sense that
only two-particle 
scattering states outside  the model space are accounted for. 
This means that both particles must have momenta larger than $k_F$ 
and that at least one of the fermions in the two-particle wave function
has momentum larger than $k_M$. 
The model space Pauli operator defined in Eq.\ (\ref{eq:pauliop})
exclude scattering states with both momenta below $k_M$. 
In the lowest order, MBHF calculations for the nuclear matter binding 
energy is performed in two steps. 
First, Eqs.\ (\ref{eq:mbhf}) and (\ref{eq:selfconx}) are
solved self-consistently.
In the second step a ``standard'' Brueckner type calculation 
(see Eq. (\ref{eq:g_bhf}) is performed
except that the MBHF spectrum  defined in Eq.\ (\ref{eq:selfcon})
is used for $k_F<k<k_M$.

\subsection{MBHF Results }
\begin{figure}[hbtp]
   \setlength{\unitlength}{1mm}
   \begin{picture}(80,115)
   \put(25,3){\epsfxsize=10cm\epsfbox{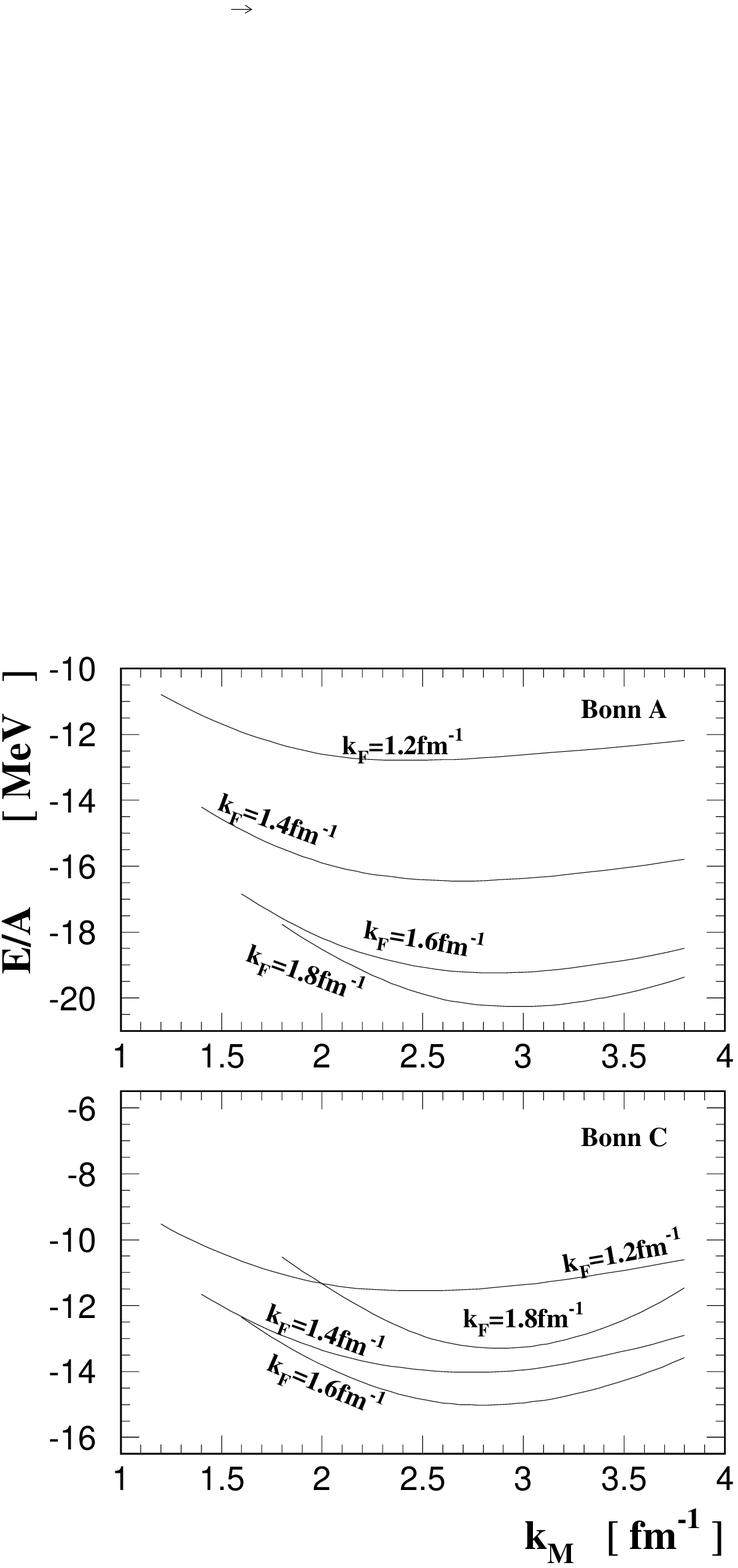}}
   \end{picture}
   \caption{ Average energy per nucleon as a function of $k_M$ 
             for MBHF calculations using version A and C of the 
             Bonn potentials. 
             The results are for four typical Fermi momenta.}
   \label{fig:mbhfkm}
\end{figure}

We have performed lowest order MBHF calculations for various 
$k_M$ values 
within the upper bound $k_M=$ 3.8 fm$^{-1}$.
In Fig.\ \ref{fig:mbhfkm} we present the  average energy per 
particle as a function of $k_M$
for three typical Fermi momenta, using  the Bonn A  and C 
potentials.  
For small Fermi momenta ($k_F=1.2$ fm$^{-1}$) the results for 
the Bonn A potential are very insensitive to the choice of $k_M$. 
For a wide region of $k_M$ (2.0-3.5 fm$^{-1}$) the energy variation 
is less than 0.5 MeV.  
For larger Fermi momenta we see that the variation is somewhat 
stronger, however, we find a "saturation" behaviour in the vicinity 
of $k_M\approx 3.0 fm^{-1}$. 
This is a fortunate and desirable result.
Here we have a minimum for $E/A$ and moreover it is here that $E/A$ 
seems to be least sensitive to $k_M$. Thus we believe that 
$k_M=$ 3.0 fm$^{-1}$ is an optimum choice for the  MBHF calculations 
using the Bonn A potential. 
Note that intermediate 
states with large momentum components ($k>$ 3 fm$^{-1}$ are 
induced by the short-range part  of the NN interaction 
\cite{km83,day67,htb70}.  Thus, using $k_M \approx$ 3.0 fm$^{-1}$
the intermediate states induced by the short-range repulsion are  
included in the model space $G$-matrix.
For the Bonn C potential  we get very similar results, although there 
is a slightly larger $k_M$-dependence than for the A potential.
For both potentials  we find that $E/A$ varies no more than 0.5 MeV 
for $k_M$ between 2.5 fm$^{-1}$ and 3.2 fm$^{-1}$. 
Similar results are also found for the Bonn B potential.
\begin{figure}[hbtp]
\setlength{\unitlength}{1mm}
\begin{picture}(80,100)
\put(25,3){\epsfxsize=12cm\epsfbox{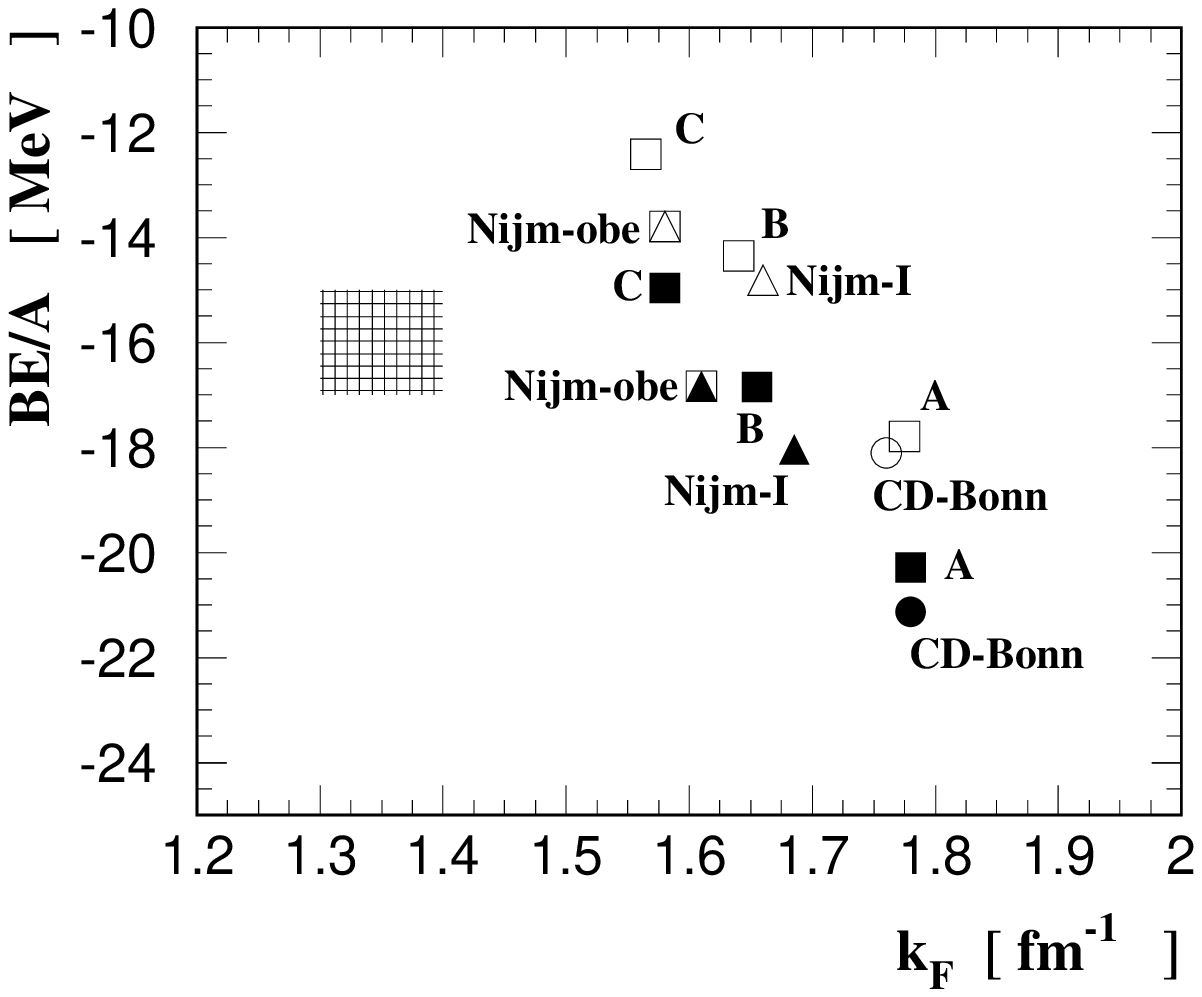}}
\end{picture}
\caption{ Nuclear matter saturation points obtained in BHF calculations
(opens symbols). MBHF calculations using $k_M=3.0$ fm$^{-1}$ are 
indicated by filled symbols. 
The results using versions A, B and C of the Bonn potential are 
shown by squares while the results for the charge dependent 
Bonn potential (CD-Bonn) are indicated  by circles. 
The triangles represent the results using Nijmegen potential 
type I (Nijm-I) and the results from the  Nijmegen meson exchange  
version  (Nijm-OBE) are denoted by triangle+square. }
\label{fig:mbhfsat}
\end{figure}

One has to keep in mind that the strengths of the tensor force of the 
potentials discussed are different, as Bonn A contains the  
weakest tensor force component of the three potentials considered, 
while Bonn C contains the strongest one.
Since all three potentials reproduce the same set of 
scattering data, a potential  containing a weaker 
tensor component needs a strong central component.
Intermediate states induced by the tensor force are relatively 
more strongly dominated by lower momentum components than states 
induced by the central force\cite{mac89}. 
Therefore, one can expect that the 
energy contribution from scattering into states with momentum 
components lower than $k_M$ will be larger for potential C than for A. 
This is also reflected in the larger $k_M$-dependence in the 
calculations using potential C. 
The results using the other potentials are very similar and we 
find that 
it is a good approximation to use the model-space boundary 
$k_M=$3.0 fm$^{-1}$ for all potentials considered.
In Fig. \ref{fig:mbhfsat} the saturation points for calculations using
$k_M=$3.0 fm$^{-1}$ are indicated  by filled symbols while the 
saturation points from standard BHF calculations, 
which correspond to $k_M=k_F$, are indicated by open symbols. 
We see that the MBHF approach yields about 3 MeV 
additional binding energy compared to the BHF approach. 
The saturation density is slightly increased. 
The results show the well-known fact that within lowest-order 
Brueckner theory one is not able to reproduce 
the empirical data for nuclear matter. In the next section the 
MBHF method is extended to include additional many-body terms, such
as the PPHH diagrams.

\section{Summation of Particle-particle and Hole-hole Diagrams 
         for Nuclear Matter}
\label{sec:pphh}
\begin{figure}[h]
       \setlength{\unitlength}{1mm}
       \begin{picture}(80,20)
       \put(25,5){\epsfxsize=5cm \epsfbox{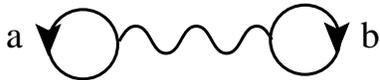}}
       \end{picture}
       \caption{First order contribution 
                to the ground-state energy shift 
                $ \Delta E_0$ in BHF theory.}
       \label{fig:gbhf}
\end{figure}

The details of the PPHH-diagram method for nuclear matter can be found
elsewhere \cite{shk87}. However, to 
facilitate the presentation and discussion
of our present work, we will give  a brief
review of the method.
In perturbative approaches to nuclear matter, 
the well-known BHF theory has been the standard starting point.
The effective two-body interaction in nuclear matter has then
been given by the so-called 
$G$-matrix, which includes all ladder-type diagrams with 
particle-particle intermediate states to infinite order. 
However, in terms of the $G$-matrix it is nevertheless only
a lowest-order theory. Other many-body contributions like
screening terms or more complicated many-body terms
are missing. The ground-state energy shift
$\Delta E_0$ in terms of the $G$-matrix is represented
by the first-order diagram  of Fig.\ \ref{fig:gbhf} and reads
\begin{equation}
        \Delta E_0^{BHF}=
	\sum _{ab} n_a n_b\langle ab\vert G^{BHF}(\omega =
        \epsilon _a + \epsilon _b )\vert ab\rangle.
        \label{eq:g_bhf}
\end{equation}
In Eq.\ (\ref{eq:g_bhf}) the n's are the unperturbed Fermi-Dirac 
distribution functions, namely $n_k$ =1 if $k\leq k_F$ and =0 
if $k>k_F$ where  $k_F$ is the Fermi momentum.
The  sp energies are denoted by $\epsilon$, 
and are determined self-consistently using the BHF theory.
\begin{figure}
       \setlength{\unitlength}{1mm}
       \begin{picture}(80,60)
       \put(25,3){\epsfxsize=10cm \epsfbox{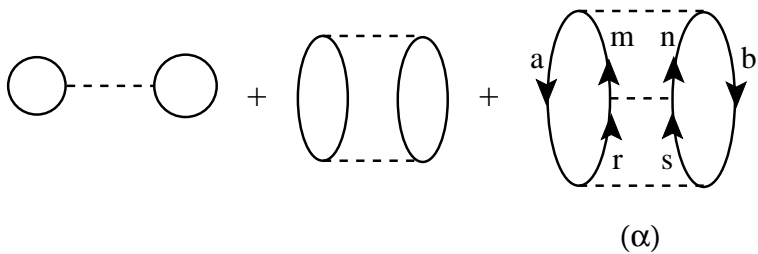}}
       \end{picture}
        \caption{Goldstone diagrams contained in the 
         Brueckner-Hartree-Fock  $G$-matrix $G^{BHF}$.}
       \label{fig:vdiagr}
\end{figure}

The Brueckner-Hartree-Fock  
$G$-matrix $G^{BHF}$ contains  repeated interactions
between a pair of "particle" lines, as illustrated by the diagrams  of 
Fig.\ \ref{fig:vdiagr}.
Note that they are so-called Goldstone diagrams, with
an explicit time ordering.
The third-order diagram $(\alpha)$ of 
Fig.\ \ref{fig:vdiagr} is given by
\begin{equation}
      Diag.(\alpha)=(\frac{1}{2})^3\frac{V_{abmn}V_{mnrs}V_{rsab}}
      {(\epsilon_a+\epsilon_b-\epsilon_m-\epsilon_n)
      (\epsilon_a+\epsilon_b-\epsilon_r-\epsilon_s)}.
\end{equation}
Here m, n, r, s are all particle lines, and $V_{ijkl}$ represents
the anti-symmetrized matrix elements of the NN interaction V.
\begin{figure}[hbtp]
       \setlength{\unitlength}{1mm}
       \begin{picture}(80,50)
       \put(40,3){\epsfxsize=3.3cm \epsfbox{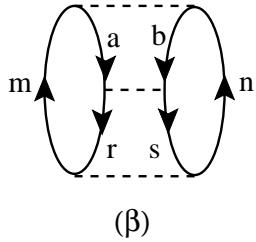}}
       \end{picture}
       \caption{Goldstone diagram with repeated 
                interactions between hole 
                lines.}
       \label{fig:vdiagrb}
\end{figure}

To the same order, there is also a diagram
with hole-hole interactions, as  shown by diagram $(\beta)$ in 
Fig.\ \ref{fig:vdiagrb}, where a, b, r, s are all hole lines.
This diagram is not included in standard BHF calculations of 
nuclear matter, for the following reason.
In earlier times, nuclear-matter calculations were based on, 
by and large, the so-called hole-line-expansion. 
The essence of the hole-line approach is that diagrams with (n+1) 
hole lines are generally much smaller than those with n hole lines.
With this criterion,  diagram ($\beta$) which has 3 (independent) 
hole lines would be negligible compared with diagram $(\alpha)$ 
which has 2 hole lines. Thus the former can be neglected. 
This approximation has, however, not been rigorously checked. 
To investigate the validity of this criterion,
it may be useful to actually calculate diagrams like ($\beta$).

\begin{figure}[hbtp]
       \setlength{\unitlength}{1mm}
       \begin{picture}(80,60)
       \put(40,5){\epsfxsize=4cm \epsfbox{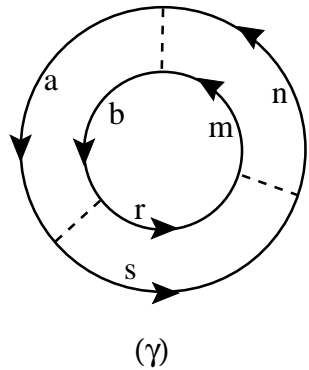}}
       \end{picture}
       \caption{Green's function diagram 
         discussed in text. The states  
         denoted by r and s can either 
         be particles or holes.}
       \label{fig:vring}
\end{figure}

A motivation behind the PPHH-diagram method of nuclear matter 
is also to include
diagrams with hole-hole correlations like diagram $(\beta)$.
In so doing, the above time-ordered formalism is no longer convenient.
We need to use the Green's function formalism, where all time variables
are integrated over the same time interval from $-\infty$ to $\infty$.
Diagrams $(\alpha )$ and $(\beta )$ are both time ordered diagrams.
In the Green's function formalism, the sum of these two diagrams
becomes a single diagram, namely diagram $(\gamma)$ of 
Fig.\  \ref{fig:vring}.  Note that we choose to draw Green's function 
diagrams differently from the time-ordered ones. 
The diagram of Fig.\ \ref{fig:vring} is given by
\begin{eqnarray}
      	Diag.(\gamma) &&=   
	\frac{-1}{2\pi i}\int_{-\infty}^{\infty}d\omega
	e^{i\omega 0^+} \nonumber \\
     	\times && {\em tr}[\frac{1}{3}F_{ab}(\omega)\bar V_{abmn} 
      	F_{mn}(\omega)\bar V_{mnrs} F_{rs}(\omega)\bar V_{rsab}] ,
      	\label{eq:3rdpphh}
\end{eqnarray}
where $F$ represents the free PPHH Green's function
\begin{eqnarray}
      F_{ab}(\omega)=\frac{(1-n_a)(1-n_b)}
      {\omega-(\epsilon_a+\epsilon_b)+i0^+}
      -\frac{n_an_b}{\omega-(\epsilon_a+\epsilon_b)-i0^+},
      \label{eq:freeg}
\end{eqnarray}
and $\bar V$ is related to V by
\begin{equation}
      \bar V_{abcd}\equiv \frac{1}{2}V_{abcd}.
      \label{eq:vdef}
\end{equation}
 In the above, the symbol $\em tr$
(trace) means that all indices are summed over.

The advantage of using this Green's function formalism is that all
the PPHH diagrams, 
of the general structure shown by Fig.\ \ref{fig:vring}, can
be expressed in a simple algebraic form. In fact the entire
PPHH-diagram series of this type becomes
\begin{eqnarray}
       "PPHH" &=& \frac{-1}{2\pi i}\int_0^1   	
       \int_{-\infty}^{\infty}d\omega
       e^{i\omega 0^+}  {\em tr}[F(\omega)\bar V \nonumber \\
       && +\frac{1}{2}(F(\omega)\bar V )^2
	+\frac{1}{3}(F(\omega)\bar V)^3 + \cdots] \nonumber \\
&=& \frac{-1}{2\pi i}\int_0^1 \frac{d\lambda}{ \lambda}  	
       \int_{-\infty}^{\infty}d\omega
       e^{i\omega 0^+}  {\em tr}[F(\omega)\bar V\lambda \nonumber \\
       && +(F(\omega)\bar V \lambda)^2
       +(F(\omega)\bar V\lambda)^3 + \cdots].
       \label{eq:ring_example}
\end{eqnarray}
Here, we have introduced the integration over the parameter 
$\lambda$, in order 
to cast the summation over all terms into a simple geometrical
series, without the factors $\frac{1}{2}$, $\frac{1}{3}$ 
and so forth.

The above expression is not yet suitable for nuclear-matter 
calculations. As is well known, the free NN interaction has a very 
strong repulsive core. The above series, expressed in terms of V, 
is  not well defined as each of its terms is very large. 
A standard way  to overcome this difficulty is to convert 
the above series into a series in the terms of the $G$-matrix.
Song {\em et al.} \cite{shk87} have shown that within the model 
space approach the PPHH-diagram series can be rewritten in terms 
of $G$-matrix vertices.
For example a third-order $G$-matrix PPHH diagram is given by
the diagram of Fig.\ \ref{fig:vring}, 
with each of its dotted-line interactions replaced by a
$G$-matrix interaction. In this way, the contribution of the entire 
PPHH-diagram series to the nuclear-matter ground-state energy can be
rewritten as
\begin{eqnarray}
      \Delta E_0^{PPHH} &=&\int ^1_0 d\lambda
      \sum _{m\in (A-2)}\sum _{abcd\in P}
      Y_{m}(ab,\lambda)Y^*_{m}(cd,\lambda) \nonumber \\
      &&\langle ab\vert G^{M}(\omega =E_m
             \vert cd\rangle ,
      \label{eq:energy}
\end{eqnarray}
where $P$ denotes the chosen model space, 
within which all nucleons are restricted
to have momentum less than $k_M$.
The  $G^M$ and $\epsilon ^M$
are the model-space $G$-matrix and sp 
spectrum defined in Eqs. (\ref{eq:mbhf}) and (\ref{eq:selfconx}),
respectively. To obtain the latter equation, a contour integration 
closed in the upper $\omega$-plane, see Ref.\ \cite{shk87} 
for further details, has been performed. 
Moreover, since by construction the model-space BHF $G$-matrix
has no poles for negative energies $\omega$, one avoids thereby 
imaginary contributions to the energy. This is perhaps the main
advantage of the model-space approach for calculating the
PPHH diagrams compared with the SCGF method of Dickhoff, Polls and 
Ramos \cite{rpd89}. However due to this approach, 
as stated previously, there are other
observables of great interest which one is not able to calculate
within the present scheme, namely
one-body Green's function, spectral
functions and momentum distributions. 

Note that the above expression is rather similar to the BHF expression
of Eq.\ (\ref{eq:g_bhf}). 
There the weighting factors are $n_an_b$ and the $G$-matrix is
evaluated at an energy ($\epsilon_a+\epsilon_b$). 
Here the weighting factors
are $YY^*$ and the $G$-matrix is evaluated at energies $E_m$.
The transition amplitudes Y and  energies $E_m$ are given
by the  RPA equation
\begin{eqnarray}
      \sum_{ef}[ (\epsilon^M _i+\epsilon^M _j)\delta_{ij,kl}
&+& (1-n_i-n_j)\lambda G^M(\omega) ] Y_m(ef,\lambda)  \nonumber \\
&&     = \ \mu_m(\omega,\lambda )Y_m(ij,\lambda),   
       \label{eq:rpaeq}
\end{eqnarray}
with the self-consistent condition
\begin{equation}
      \omega=\mu_m(\omega,\lambda)\equiv E_m.
      \label{eq:selfcons}
\end{equation}
For notational economy we will write $\mu_m(\omega,\lambda)=\mu_m$.

It is important to mention that the normalization of the RPA wave 
functions $Y$ is given by
\begin{equation}
\sum _{ab}(1-n_a-n_b)Y_m(ab,\lambda)Y_m^*(ab,\lambda)
  =\frac{-1}{1-\frac{\partial \mu_m}{\partial \omega}
   \vert _{\omega=\mu _m}}
   \label{eq:rpanorm}
\end{equation}
where  $\mu_m\equiv\mu_m(\omega,\lambda)$.
For $\lambda =1, $ $E_m$ and $Y_m$ of Eq.\ (\ref{eq:rpaeq}) are both
physical quantities \cite{wukuo}, namely
\begin{equation}
  E_m=
 (E^{A}_{0}-E^{A-2}_{m}),
\end{equation}
and
\begin{equation}
Y_m(\alpha \beta)\equiv \langle \Psi^A_0
\vert a^+_{\alpha}a^+_{\beta}
\vert  \Psi ^{A-2}_m \rangle
\end{equation}
where $E^A_0$ and $\Psi ^A_0$ represent the energy and wave 
function of the nuclear matter ground state, and similarly 
$E^{A-2}_m$  and $\Psi ^{A-2}_m$  the m'th state of the 
nuclear matter system with two less nucleons.

\begin{figure}[hbtp]
       \setlength{\unitlength}{1mm}
       \begin{picture}(70,60)
       \put(25,5){\epsfxsize=10cm \epsfbox{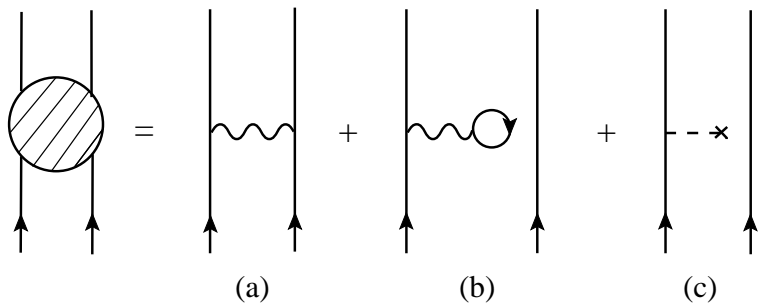}}
       \end{picture}
       \caption{Structure of PPHH vertex function $\Gamma$ 
                used to calculate PPHH diagrams in this work.
                The two-body part is the model space $G$-matrix 
		$G^M$ (a). 
                Diagram (b) is the one-body term from  $G^M$ and 
                diagram (c) is the one-body potential introduced 
                $u^M$ in the  Hamiltonian.}
       \label{fig:vertex}
\end{figure}

Eq.\ (\ref{eq:rpaeq}) 
is a standard RPA secular equation for the PPHH Green's function, and
as shown  its irreducible vertex function  is given by  the two-body
 model-space $G$-matrix $G^M$ only. This is, however, inadequate
in order to  describe the Green's function, as there are also 
one-body terms in the PPHH vertex function. Such one-body diagrams 
are important for the PPHH Green's functions
(see ref.\ \cite{wukuo} and references quoted therein). 
To first order in $G^M$,
we should also include in the vertex function one-body diagrams. 
The resulting vertex function, denoted by $\Gamma$, is then composed 
of the three diagrams of Fig.\ \ref{fig:vertex}. 
Here diagram (a) is the two-body $G^M$ term, while diagram
(b) is the one-body $G^M$ term. We have added and then subtracted 
a one-body sp potential $u^M$ to the Hamiltonian, namely 
$H=T+V=(T+u^M)+(V-u^M) \equiv H_0+H_I$. 
Diagram (c) of Fig.\ \ref{fig:vertex} is the $-u^M$ term.
\begin{figure}[hbtp]
       \setlength{\unitlength}{1mm}
       \begin{picture}(60,50)
       \put(40,5){\epsfxsize=4cm \epsfbox{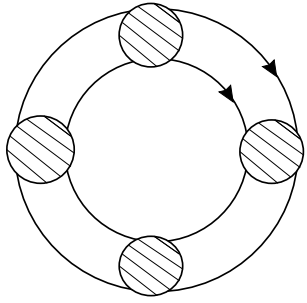}}
       \end{picture}
       \caption{An example of PPHH diagrams 
              included in our calculations.}
       \label{fig:pphhdiag}
\end{figure}

The above concern about the vertex function suggests that for nuclear
matter calculations we should
consider a generalized PPHH-diagram series, as represented by 
Fig.\ \ref{fig:pphhdiag}.
There each "blob" represents the vertex function $\Gamma$
of Fig.\ \ref{fig:vertex}. 
In this way the nuclear-matter PPHH diagrams include not only
two-body interactions
via $G_M$ but also one-body insertions.  The all-order sum of these 
generalized PPHH diagrams can still be readily performed, with the
final result  still given by Eq.\ (\ref{fig:pphhdiag}), except that
the amplitudes $Y$ and energies $E$ are now given by a modified RPA 
equation, namely
\begin{eqnarray}
   \sum_{ef}[(\epsilon^M _i  + \epsilon^M _j)\delta_{ij,ef}
   &+& (1-n_i-n_j)\lambda \Gamma(\omega) ]Y_m(ef,\lambda) \nonumber \\
   &&=\ \mu_m(\omega,\lambda)Y_m(ij,\lambda).
   \label{eq:rpa}
\end{eqnarray}
A computational subtlety may be mentioned. The above RPA equation 
are to be solved  with the self-consistent condition 
(\ref{eq:selfcons}). 
It is rather complicated to do so
numerically, and it was mainly because of this consideration that
earlier calculations \cite{shk87} treated this self consistency
by way of a perturbation method. In the present work we have 
treated this
self-consistency exactly.

The inclusion of the one-body insertions  for the PPHH diagrams
has played an important role in our calculation. When such insertions
are included,  the wound-integral effect is present in the PPHH-diagram
calculation. Because of the interaction among nucleons, particularly
the short range correlation among them, nucleons in the model space are
part of the time outside the model
space. This "absence" from the model space is represented by the wound
integral. At low density, the wound integral is generally small
and the normalization factor in Eq.\  (\ref{eq:rpanorm}) of the RPA 
amplitudes is close to -1.
As the density increases, the wound-integral  becomes larger
and the normalization can become much smaller, such as -0.8.
As a result, the PPHH-diagram contribution to the nuclear 
matter potential
energy, as given by Eq.\ (\ref{eq:energy}), is suppressed as 
density increases.
In short,
the inclusion of the one-body insertions has been essential  for the
determination of the nuclear matter saturation density, as has been
observed in ref.\cite{shk87}.
 
As mentioned earlier, the PPHH diagrams are calculated within the 
model space P, where all nucleons are restricted to have momentum 
up to $k_M$. A main concern has been what value for $k_M$ one should 
use. In section \ref{sec:mbhf} we have pointed out that $k_M$ should 
not exceed 3.8 fm$^{-1}$. For $k_M = k_F$, the PPHH-diagram approach 
reduces to the standard BHF method.
Thus the range for $k_M$ is between $k_F$ and 3.8 fm$^{-1}$.

\section{Results of PPHH-diagram Calculations}
\label{sec:result}

\begin{figure}[hbtp]
     \setlength{\unitlength}{1mm}
     \begin{picture}(60,80)
      \put(25,0){\epsfxsize=12.0cm \epsfbox{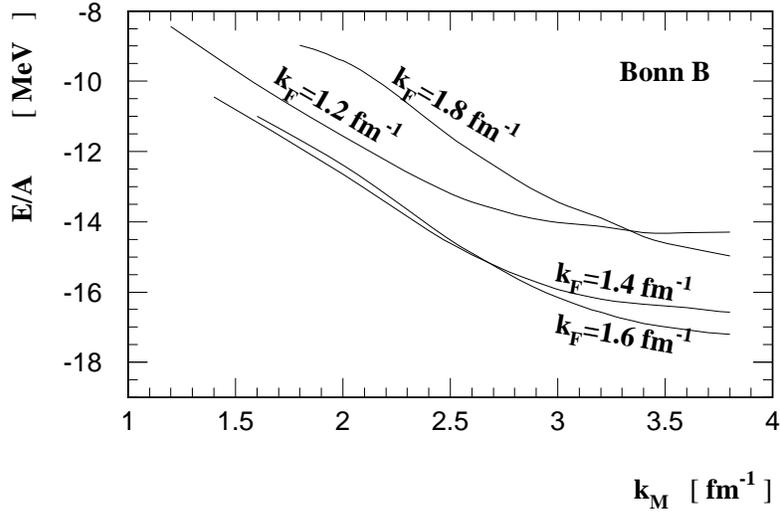}}
     \end{picture}
     \caption{Results for the version B of the Bonn potential. 
              See Fig.\ 9 for further explanations.}
       \label{fig:bring}
\end{figure}

\begin{figure}[hbtp]
     \setlength{\unitlength}{1mm}
     \begin{picture}(60,80)
     \put(25,0){\epsfxsize=12.0cm \epsfbox{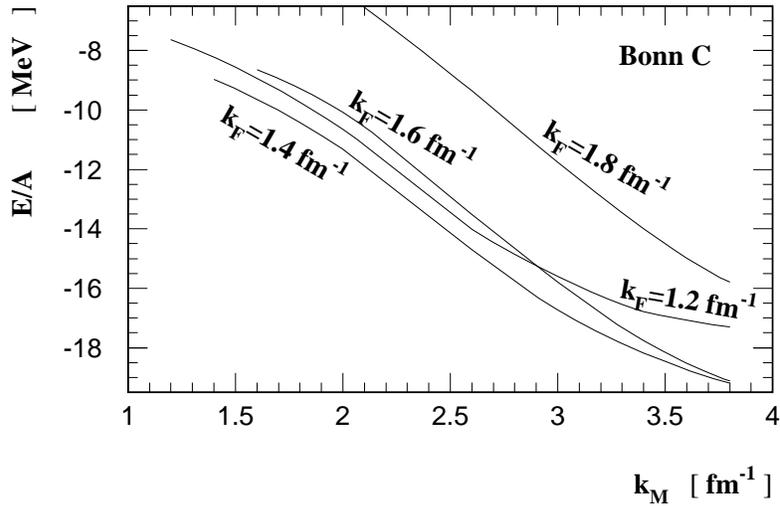}}
     \end{picture}
     \caption{Results for the version C of the Bonn potential.
              See Fig.\ 9 for further explanations.}
       \label{fig:cring}
\end{figure}

We have performed PPHH-diagram nuclear matter calculations
for a range of $k_M$ values within the above upper bound. For
each $k_M$ value, we first calculate the MBHF sp 
spectrum for various
Fermi momenta ($k_F$). Then we solve the RPA equation in 
Eq.\ (\ref{eq:rpaeq})  for several values of $\lambda$. 
The resulting wave functions $Y$ and eigenvalues $E$ are then used
to evaluate the PPHH-diagram sum in Eq.\ (\ref{eq:energy}).
\begin{figure}[hbtp]
      \setlength{\unitlength}{1mm}
      \begin{picture}(60,80)
      \setlength{\unitlength}{1mm}
      \put(25,0){\epsfxsize=12.0cm \epsfbox{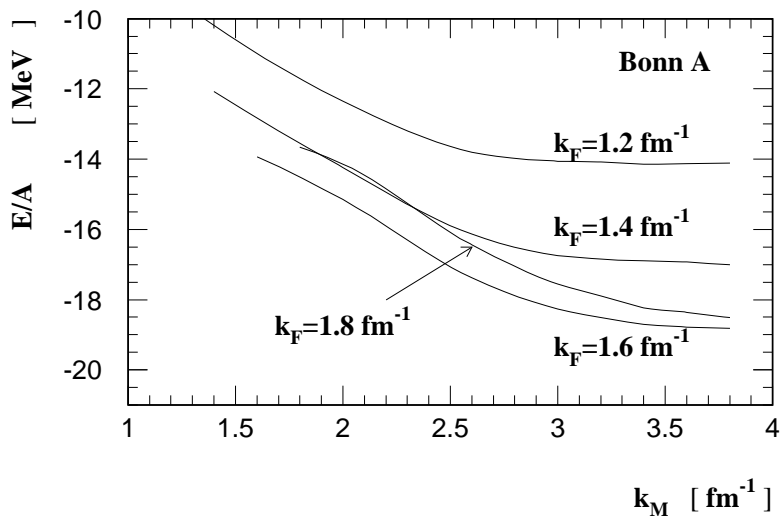}}
      \end{picture}
      \caption{The average energy per nucleon as a 
               function of $k_M$ for fixed values of $k_F$.
               The results are obtained with the version A of the 
               Bonn potential.}
       \label{fig:aring}
\end{figure}
In Fig.\ \ref{fig:aring} the average energy per 
particle $E/A$ is shown as a function of $k_M$ for four typical 
Fermi momenta using version A of the Bonn potential. The results 
are quite stable with variations in $k_M$ in the vicinity of 
$k_M$ = 3.8 fm$^{-1}$. The corresponding results using 
the Bonn B and C potentials are shown 
in Figs.\ \ref{fig:bring} and  \ref{fig:cring}, respectively.
We see that the results using  Bonn A and B are very similar, 
suggesting that $k_M$ = 3.8 fm$^{-1}$ is an optimal choice 
for calculating PPHH diagrams.
However, for the Bonn C potential saturation  
occurs at higher $k_M$ values.

\begin{figure}[hbtp]
     \setlength{\unitlength}{1mm}
     \begin{picture}(60,80)
     \put(25,0){\epsfxsize=12cm \epsfbox{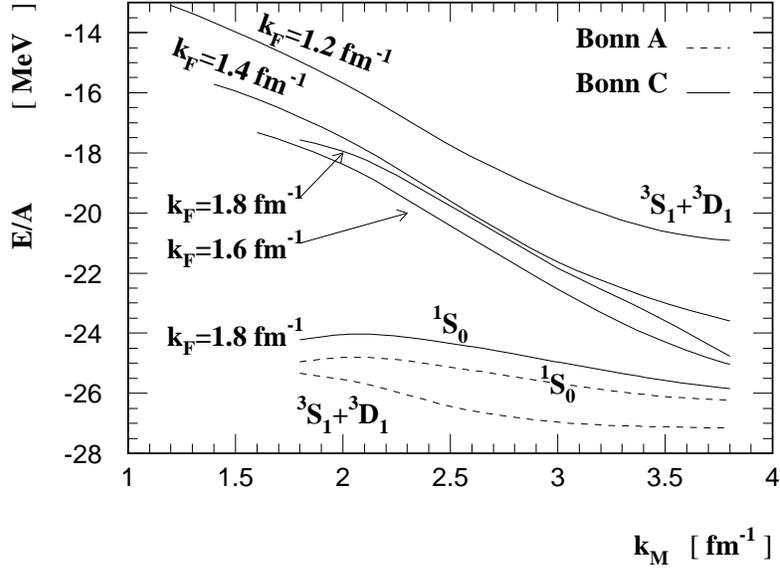}}
     \end{picture}
     \caption{Contributions to the average energy per particle 
from different partial waves.
The results are obtained with the Bonn A and C potentials.
              See Fig.\ \ref{fig:aring} for further explanations.}
       \label{fig:3s1}
\end{figure}

To understand this difference  
it may be instructive to investigate contributions from the different 
partial waves.  
In Fig.\ \ref{fig:3s1} the contributions from the $^3$S$_1+^3$D$_1$
channel are shown. Results obtained with the Bonn C potential 
are shown by solid lines. 
Comparing these
curves with those in Fig.\ \ref{fig:cring}, we see that most of the
energy gained when going from  a low to  a high $k_M$ value comes 
from this channel. Note that in this channel the tensor force 
plays an essential role. 

In the $^1$S$_0$ channel, where  the tensor force does not contribute,
the energy shift is found to be much smaller.
In Fig.\ \ref{fig:3s1} the $^1$S$_0$ contribution is shown for  
$k_F$ = 1.8 fm$^{-1}$
For  the other Fermi momenta this channel yields similar results. 
The $k_M$-dependence of the $^1$S$_0$ channel arises mainly 
in the normalization factors of Eq.\ (\ref{eq:rpanorm}).
Thus, the $k_M$-dependence in  the $^1$S$_0$ channel 
can be explained mainly by the $\omega$-dependence of the vertex 
function $\Gamma$ 
in RPA equation (\ref{eq:rpa}).
Recall that $\omega$-dependent one-body terms are included in $\Gamma$.
When the RPA equation\ (\ref{eq:rpa}) is decomposed into separate 
partial wave channels, 
the one-body terms for the various channels are equal.

For higher order partial waves the (e.g., $l\geq 4$)
$\omega$-dependence of the $G$-matrix is known to be very weak and
the $G$-matrix differs little from the bare interaction.
Therefore, the $k_M$-dependence for all higher order partial waves is 
mainly due to the $\omega$ dependent one-body terms and shows up 
in the normalization factor. 

For comparison we have also included the $^3$S$_1+^3$D$_1$ and 
$^1$S$_0$ contribution for potential A.
We see that the $^3$S$_1+^3$D$_1$  
channel for this potential  has a much weaker $k_M$-dependence. 
Note that the Bonn A potential has a weaker tensor force than the 
Bonn C potential.  
This indicates that the strength of the tensor force is important 
for the $k_M$-dependence of the calculations.   
In the $^1$S$_0$ channel potentials A and C yield similar results.
The small difference can be explained by the
one-body terms of potential C which generally 
have a stronger $\omega$-dependence than those of potential A.
This is reflected in different normalization factors for the two
potentials.
  
\begin{figure}[hbtp]
     \setlength{\unitlength}{1mm}
     \begin{picture}(60,190)
     \put(25,120){\epsfxsize=12cm \epsfbox{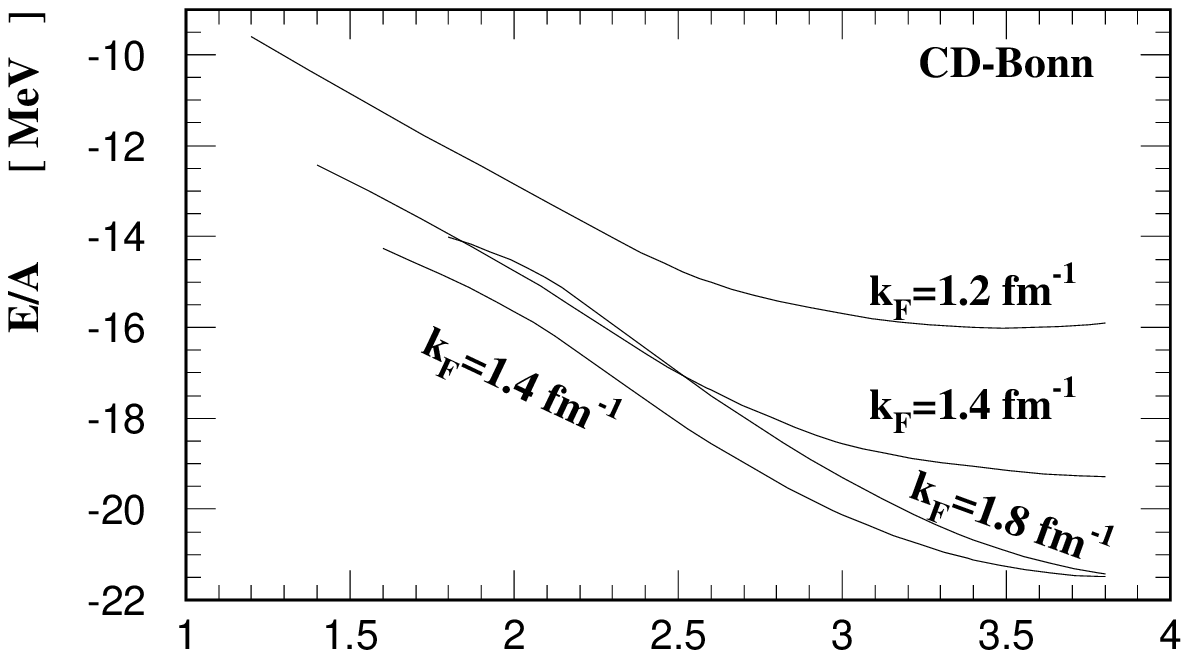}}
     \put(25,60){\epsfxsize=12cm \epsfbox{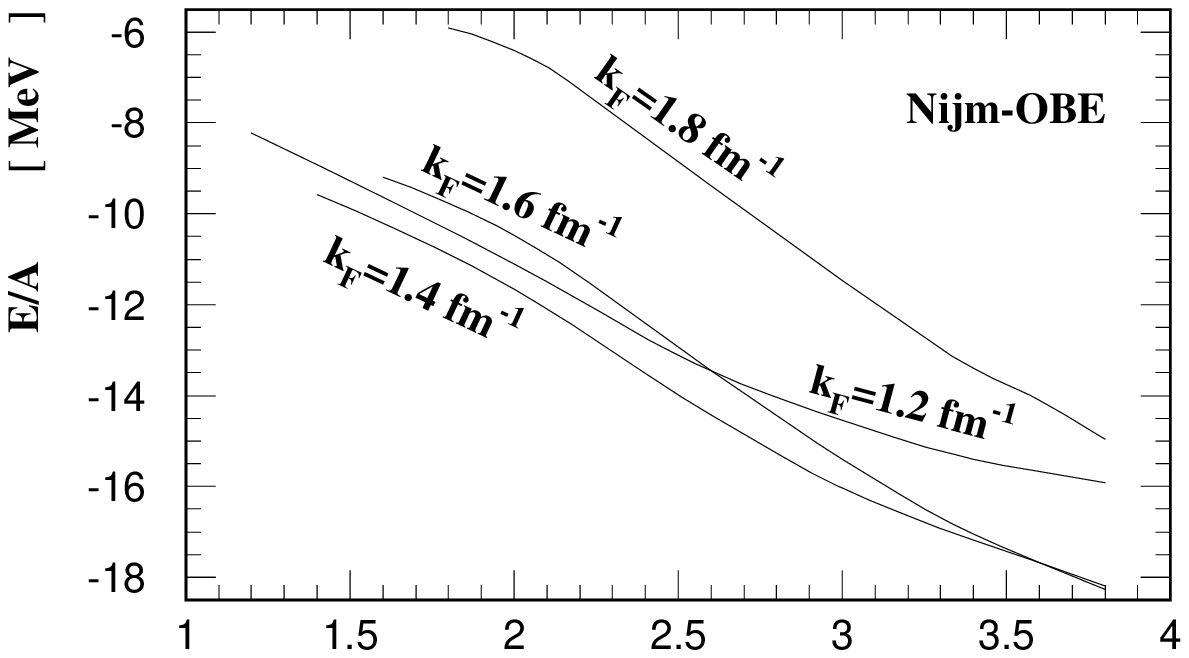}}
     \put(25,0){\epsfxsize=12cm \epsfbox{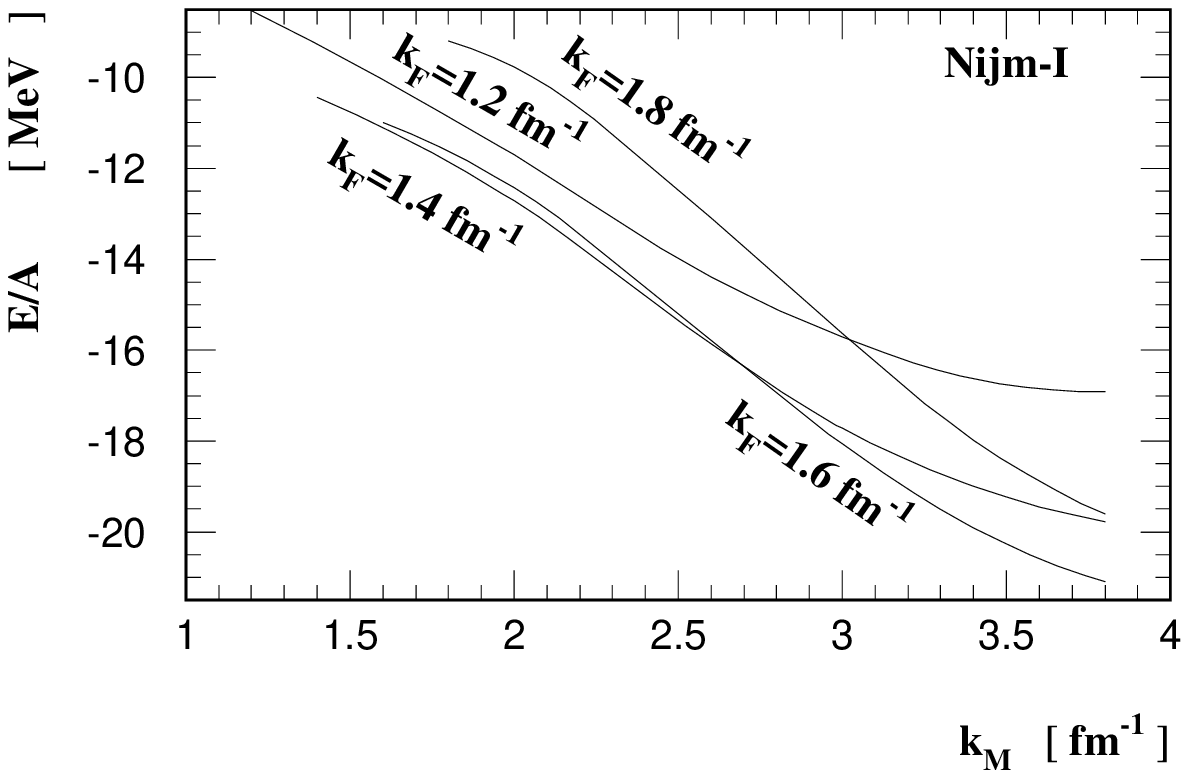}}
     \end{picture}
      \caption{The average energy per nucleon as a 
               function of $k_M$ for fixed values of $k_F$.
		Results are obtained with the CD-Bonn, 	
		Nijm-OBE and Nijm-I 
		potentials.}
       \label{fig:kmnewv}
\end{figure}

To complete our results we have included results  for the 
CD-bonn potential \cite{cdbonn} and two versions of the 
Nijmegen group \cite{nim}, the
 meson-exchange potential (Nijm-OBE) and a local potential with 
a non-local contribution to the central force, i.e.\ the model
Nijm-I of Ref.\ \cite{nim}.The results are shown in 
Fig. \ref{fig:kmnewv}. 
We see that the results are similar to those using the earlier 
versions of the Bonn potential, in the sense that when the strength 
of the tensor force increases, 
the stability of the results as functions of $k_M$ worsens.

We find that the BHF calculations  for Bonn B and Nijm-I yields
similar results. Note, however, that the tensor force is stronger 
for the Nijm-I potential. The D-state probability of the deuteron
reflects the tensor force in the NN-potential. The prediction from 
the Nijm-I potential  is 5.66 \%  while the Bonn B and C potentials 
yields 5.0 \%\ and 5.6 \%\ respectively.
Thus, the tensor force for the Nijm-I may be  more comparable 
with that of the Bonn C potential.  
This is also reflected in the large PPHH-diagram 
contribution for the Nijm-I potential. 

\begin{figure}[hbtp]
\setlength{\unitlength}{1mm}
\begin{picture}(60,90)
\put(25,3){\epsfxsize=12cm\epsfbox{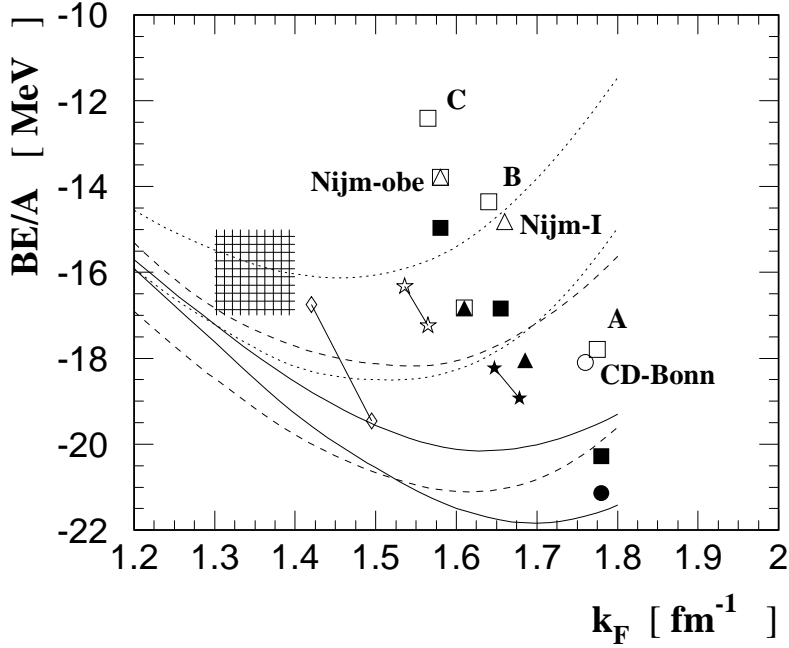}}
\end{picture}
\caption{ Nuclear matter saturation curves (PPHH-diagram calculations) 
and points obtained with various potentials.
The solid curves are obtained with the CD-Bonn potential, while  
dotted and dashed curves are obtained with the Nijm-OBE and Nijm-I 
respectively. 
For each potential calculations using $k_M$ = 3.0 fm$^{-1}$  
(upper curve) and $k_M$ = 3.8 fm$^{-1}$ (lower curve) are presented.
For the Bonn A, B and C potentials the saturation points (filled star,
open star and diamond respectively) are shown  
for the two $k_M$ values considered.
In addition the MBHF results are presented as in Fig.\ 3. }
\label{fig:pphhsat}
\end{figure}

In Fig.\ \ref{fig:pphhsat} we show the results of our PPHH-diagram 
calculation using $k_M$ = 3.0 fm$^{-1}$ and   3.8 fm$^{-1}$. 
As seen from the figure the saturation density increases slightly when
$k_M$ increases to 3.8 fm$^{-1}$. 
However, compared to the MBHF results all our PPHH-diagram 
calculations yield a lower saturation density.
Previous PPHH-diagram calculations for nuclear matter were performed
using $k_M$ =3.2 fm$^{-1}$ \cite{shk87,jia88}. 
We find that our calculations 
using the same $k_M$ value yield similar results.
Note that in this work the normalization of the RPA wave function is
performed by  the use of Eq.\ (\ref{eq:rpanorm}).
In earlier calculations\cite{shk87} a perturbation method has been  
used in order  to normalize the  RPA wave functions.

\section{Calculations for Neutron Matter}
\label{sec:neutron}
We have found that the results for the PPHH-diagram calculations are 
very sensitive to the strength of the tensor force. 
In symmetric nuclear matter the contribution from the tensor force 
to the binding energy comes mainly from the $^3$S$_1+^3$D$_1$  
channel which has isospin equal to zero. 
For neutron matter, however, this channel does not contribute,
and therefore the tensor force is not so important. 
To complete our  discussion of the PPHH diagrams we therefore
include results for neutron matter as well.
\begin{figure}[hbtp]
\setlength{\unitlength}{1mm}
\begin{picture}(60,100)
\put(25,3){\epsfxsize=12cm\epsfbox{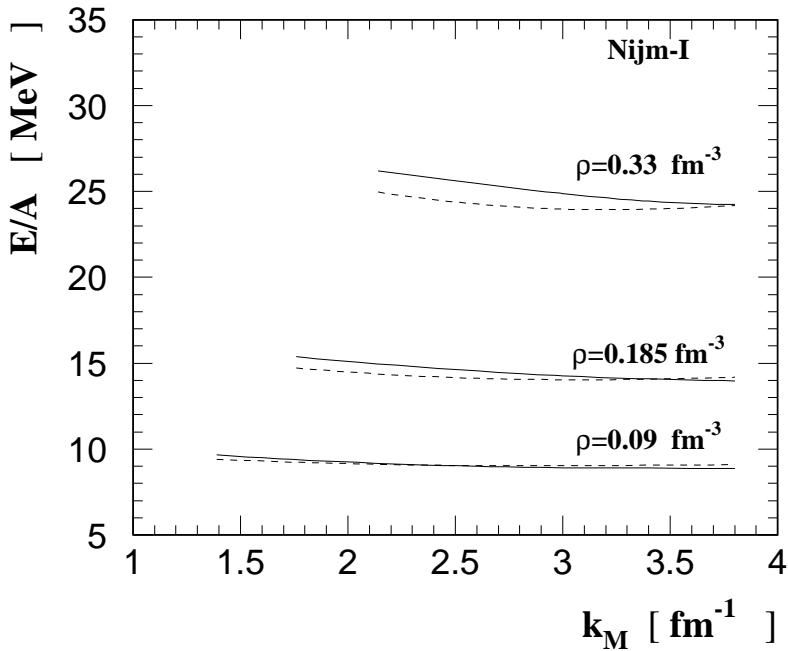}}
\end{picture}
\caption{ The average energy per neutron for neutron matter  
as a function of $k_M$ for various densities. The MBHF results are 
shown by dashed lines while the results including PPHH diagrams 
are shown by solid lines.} 
\label{fig:kmneutr}
\end{figure}
In Fig.\ \ref{fig:kmneutr} we show the average energy per nucleon as a 
function of $k_M$ for various densities. In this figure we show only
results for the Nijm-I potential, since the other potentials
gave qualitatively similar results. The results from the PPHH-diagram
calculation are given  
by solid lines. For the MBHF results, which are shown by dashed lines, 
the energy does not shift more than about 1 MeV if 
$k_M$ increased from its lowest value ($k_M=k_F$) to 
$\approx$ 3.0 fm$^{-1}$.
We see that the $k_M$-dependence is very weak in both
the MBHF calculations and the calculations including PPHH diagrams. 
Moreover, the contribution from PPHH diagrams are substantially reduced
compared to the results for symmetric nuclear matter. 

\begin{figure}[hbtp]
\setlength{\unitlength}{1mm}
\begin{picture}(60,90)
\put(25,3){\epsfxsize=12cm\epsfbox{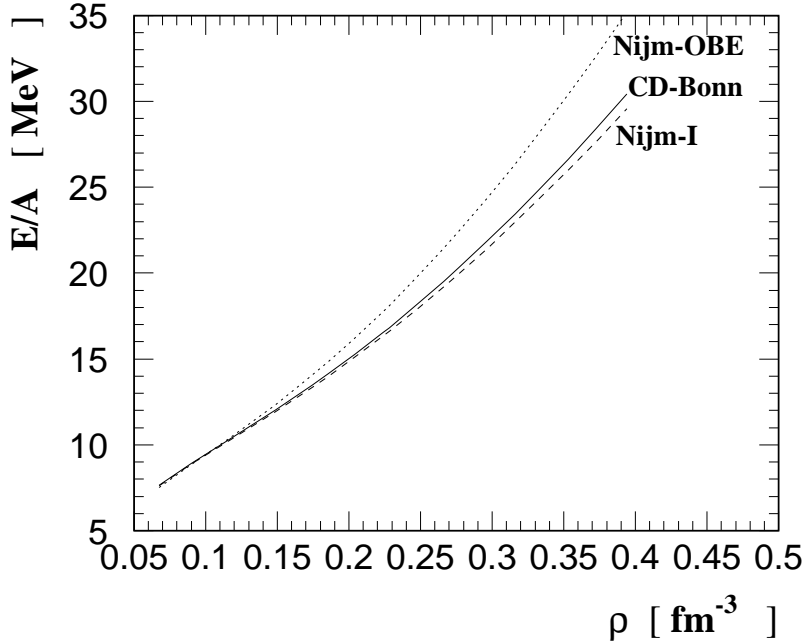}}
\end{picture}
\caption{ The average energy per neutron for neutron matter using
various potentials. PPHH diagrams are included using 
$k_M$ = 3.8 fm$^{-1}$.
The results obtained with the CD-Bonn potential are shown by the 
solid line
while the results using Nijm-OBE and Nijm-I potentials are shown by 
dotted and dashed lines respectively. }
\label{fig:neutronpphh}
\end{figure}

In Fig.\ \ref{fig:neutronpphh} the average energy is plotted as 
a function of density for the three potentials. 
Results including PPHH diagrams are shown.
However, we find that the contribution from the PPHH diagrams is small.
This has important consequences for studies of neutron star 
properties. 
Many-body terms such as hole-hole terms play a smaller role. 
It is therefore
of interest to perform a self-consistent calculation of 
particle-hole diagrams, of importance for long-range effects, in order
to assess the relation between particle-particle, hole-hole and 
particle-hole
terms in neutron matter or $\beta$-stable matter calculations. 

\section{Summary and Conclusion}
\label{sec:concl}
A central ingredient in the PPHH-diagram calculation for nuclear 
matter is its model space size $k_M$. 
How to determine it has been a long standing
problem. We have carried out extensive 
nuclear matter calculations, with various choices for $k_M$
for both the model-space Brueckner-Hartree (MBHF) approach 
and the approach which includes 
particle-particle and hole-hole (PPHH) diagrams.
We have found that for the MBHF approach  
the nuclear matter binding energy and saturation density
vary little with $k_M$ for $k_M$ in the vicinity  3.0 fm$^{-1}$.
The location of this saturation region for $k_M$ is 
almost independent of the potential used. 

For calculations  including PPHH diagrams we find a saturation 
region where the nuclear matter binding energy and 
saturation density vary rather little with $k_M$ 
for $k_M\approx$ 3.8 fm$^{-1}$ .
However, the  calculations  including PPHH diagrams exhibit a stronger 
$k_M$-dependence than the MBHF calculations.
Our results  have shown that the strength of the tensor force is 
important for the $k_M$-dependence. Potentials with weak tensor force 
yield better convergence  for large $k_M$ than those with
stronger tensor force.

The new potentials which have a $\chi^2$ fit per datum close 
to $1$ generally yield similar results as the older versions 
of the Bonn potentials. 
For all potentials considered, the saturation density
is substantially reduced when PPHH diagrams are included.

MBHF calculations for neutron matter are found to be insensitive
to the choice of $k_M$. Moreover, the contribution from PPHH 
diagrams is found to be small in this case. 
Thus, combining this  with 
our result for symmetric matter, we have shown that the contributions
from PPHH excitations in a nuclear medium depend strongly on the 
tensor force component of the NN interaction. 

We have carried out our PPHH-diagram calculations using the 
MBHF sp spectrum.
It should be of interest to employ other sp spectra in the summation 
of PPHH diagrams for 
nuclear matter calculations. As an example, it would be desirable to 
employ a sp spectrum which is obtained in a consistent way from the
PPHH-diagram calculation of the binding energy. 
As far as we know, how to formulate
such a sp spectrum is not yet fully understood. 
Such a formulation is presently being studied by us.
\begin{ack}
This work has been supported 
by The Research Council of Norway (NFR) under the Programme for Supercomputing 
through a grant of computing time.
\end{ack}


\end{document}